\documentclass[fleqn,usenatbib, letters]{mnras}
\hypersetup{draft}

\usepackage[T1]{fontenc}
\usepackage{ae,aecompl}

\usepackage{graphicx}	
\usepackage{amsmath}	
\usepackage{amssymb}	
\usepackage[french]{babel}
\usepackage[utf8]{inputenc}

\newcommand{\ha}{\ifmmode {\rm H}\alpha \else H$\alpha$\fi}
\newcommand{\hb}{\ifmmode {\rm H}\beta \else H$\beta$\fi}
\newcommand{\lya}{\ifmmode {\rm Ly}\alpha \else Ly$\alpha$\fi}

\newcommand{\vpeak}{$V_{\rm peak}^{\rm red}$}
\newcommand{\sep}{$\Delta V_{1/2}$}
\newcommand{\udft}{\textsf{udf-10}}
\newcommand{\mosaic}{\textsf{udf mosaic}}

\def\kms{km s$^{-1}$}

\def\oiii{[\ion{O}{iii}]}

\def\oiiill{[\ion{O}{iii}]$\lambda\lambda 4959,5007$}
\def\ciii{\ion{C}{iii}]}

\def\ciiill{\ion{C}{iii}]$\lambda\lambda 1907,1909$}

\title[z$_{\rm sys}$ from \lya]{Recovering the systemic redshift of galaxies from their Lyman-alpha line profile}

\author[Verhamme et al.]{A. Verhamme$^{1,2}$\thanks{E-mail: anne.verhamme@unige.ch},
T. Garel$^{1}$,
E. Ventou$^{3}$,
T. Contini$^{3}$,
N. Bouch\'e$^{3}$,
E.C. Herenz$^{14}$,
\newauthor
J. Richard$^{1}$,
R. Bacon$^{1}$,
K.B. Schmidt$^{4}$,
M. Maseda$^{5}$,
R.A. Marino$^{7}$,
J. Brinchmann$^{5,6}$,
\newauthor
S. Cantalupo$^{7}$,
J.Caruana$^{8,9}$,
B. Cl\'ement$^{1}$,
C. Diener$^{13,4}$, 
A.B. Drake$^{1}$,
T. Hashimoto$^{1,10,11}$,
\newauthor
H. Inami$^{1}$,
J. Kerutt$^{4}$,
W. Kollatschny$^{12}$,
F. Leclercq$^{1}$,
V. Patr\'icio$^{1}$,
J. Schaye$^{5}$,
\newauthor
L. Wisotzki$^{4}$,
J. Zabl$^{3}$ \\\\
$^{1}$Univ Lyon, Univ Lyon1, Ens de Lyon, CNRS, Centre de Recherche Astrophysique de Lyon UMR5574, F-69230, Saint-Genis-Laval, France \\
$^{2}$Observatoire de Gen\`eve, Universit\'e de Gen\`eve, 51 Ch. des Maillettes, 1290 Versoix, Switzerland \\
$^{3}$Institut de Recherche en Astrophysique et Planétologie (IRAP), Université de Toulouse, CNRS, UPS, F-31400 Toulouse, France \\
$^{4}$Leibniz-Institut f\"ur Astrophysik Potsdam (AIP), An der Sternwarte 16, D-14482, Potsdam, Germany\\
$^{5}$Leiden Observatory, Leiden University, NL-2300 RA Leiden, Netherlands\\
$^{6}$Instituto de Astrof{\'\i}sica e Ci{\^e}ncias do Espaço, Universidade do Porto, CAUP, Rua das Estrelas, PT4150-762 Porto, Portugal\\
$^{7}$Department of Physics, ETH Z$\ddot{u}$rich,Wolfgang$-$Pauli$-$Strasse\,27, 8093\,Z$\ddot{u}$rich, Switzerland\\
$^{8}$Department of Physics, University of Malta, Msida MSD 2080, Malta\\
$^{9}$Institute for Space Sciences and Astronomy, University of Malta, Msida MSD 2080, Malta\\
$^{10}$National Astronomical Observatory of Japan, 2-21-1 Osawa, Mitaka, Tokyo 181-8588, Japan\\
$^{11}$College of General Education, Osaka Sangyo University, 3-1-1 Nakagaito, Daito, Osaka 574-8530, Japan\\
$^{12}$Institut f\"ur Astrophysik, Universit\"at G\"ottingen, Friedrich-Hund Platz 1, D-37077 G\"ottingen, Germany\\
$^{13}$Institute of Astronomy, Madingley Road Cambridge, CB3 0HA, UK\\
$^{14}$Department of Astronomy, Stockholm University, AlbaNova University Centre, SE-106 91, Stockholm, Sweden\\
}

\date{Accepted XXX. Received YYY; in original form ZZZ}

\pubyear{2018}

\begin{document}
\label{firstpage}
\pagerange{\pageref{firstpage}--\pageref{lastpage}}
\maketitle

\begin{abstract}
The Lyman alpha (\lya) line of Hydrogen is a prominent feature in the spectra of star-forming galaxies, usually
redshifted by a few hundreds of \kms\ compared to the systemic redshift. This large offset hampers follow-up surveys, galaxy pair statistics and correlations with quasar absorption lines when only \lya\ is available. We propose diagnostics that can be used to recover the systemic redshift directly from the properties of the \lya\ line profile. We use spectroscopic observations of Lyman-Alpha Emitters (LAEs) for which a precise measurement of the systemic redshift is available. Our sample contains 13 sources detected between $z\approx3$ and $z\approx6$ as part of various  Multi Unit Spectroscopic Explorer (MUSE) Guaranteed Time Observations (GTO). We also include a compilation of spectroscopic \lya\ data from the literature spanning a wide redshift range ($z\approx0-8$). 
First, restricting our analysis to double-peaked \lya\ spectra, we find a tight correlation between the velocity offset of the red peak with respect to the systemic redshift, \vpeak, and the separation of the peaks.
Secondly, we find a correlation between \vpeak\ and the full width at half maximum of the \lya\ line. 
Fitting formulas, to estimate systemic redshifts of galaxies with an accuracy of $\leq 100$ \kms\ when only the \lya\ emission line is available, are given for the two methods.
\end{abstract}




\section{Introduction}

In the last few decades, large samples of high-redshift galaxies ($z>2$) have been assembled from deep photometric surveys based on broad/narrow-band selection techniques \citep[][and references therein]{Steidel2003,ouchi2008,Bouwens2015,Finkelstein2015,Sobral2017}. In parallel, blind spectroscopic searches commonly rely on the \lya\ line redshifted to the optical or the near-infrared to identify or confirm sources at $z\geq2$ \citep[e.g.][]{blanc2011,bielby2011,lefevre2015}. The number of spectroscopic detections of Lyman Alpha Emitters (LAE) is now being increased dramatically with ongoing observational campaigns with the Multi Unit Spectroscopic Explorer \citep[MUSE,][]{Bacon10} on ESO's VLT, allowing us to study galaxy formation and evolution with a homogeneous sample of sources over a large redshift range \citep[$2.8 \lesssim z \lesssim 6.7$, e.g.][]{Bacon15, Bacon17, Drake17, Herenz17, Mahler17, Caruana18}.

Several studies have demonstrated that the \lya\ emission line is not exactly tracing systemic redshift \citep[e.g.][]{shapley03,rakic11,mclinden2011,Song14,Hashimoto15}. Instead, the line profiles often show a complex structure which arguably originates from the propagation of resonant \lya\ photons in neutral gas within the interstellar medium and/or in the vicinity of galaxies. 
Among the broad diversity of \lya\ profiles in \lya\ emitting galaxies, we identify the most common two categories:
(i) spectra with a redshifted single peak ($\sim 2/3$ of \lya\ emitting Lyman Break Galaxies, called LBGs, from \cite{Kulas12}), and (ii) double-peaked profiles, with a prominent red peak and a smaller blue bump ($\sim 2/3$ of the remaining $1/3$ of \lya\ emitting LBGs which are multiple peaked, from \cite{Kulas12}; 40\% of the LAEs observed by \cite{Yamada12}). We will refer to the latter as \emph{blue bump LAEs} in the remainder of this paper. Understanding the nature of blue bump LAEs and studying their occurrence and evolution with redshift will be the goal of a forthcoming study. The vast majority of objects display a red peak shifted by a variable amount peaking around $\sim400$ \kms\ for Lyman Break Galaxies \citep[LBGs, e.g.][]{shapley03,Kulas12}, $\sim200$ \kms\ for LAEs \citep[LAEs, e.g.][]{Hashimoto13, Song14, Erb14, Trainor15, Henry15, Hashimoto15}, and less than $\sim150$ \kms\ for a small sample of 5 local Lyman Continuum Emitters \citep{Verhamme17}.

If not accounted for, this offset with respect to the systemic redshift can be problematic when addressing astrophysical issues which require accurate systemic redshift measurements (e.g. galaxy interactions, gas kinematics, baryonic acoustic oscillations, IGM-galaxy emission/absorption correlations).
The scope of this paper is to investigate whether the \lya\ profile shape can be used to determine the systemic redshift of galaxies. The outline of this Letter is as follows:
in Sect.2, we gather recent spectroscopic data from MUSE GTO surveys and from the literature that have sufficient spectral resolution to investigate the \lya\ line properties, as well as reliable systemic redshift measurements. In Sect. 3, We present two diagnostics which can be used to recover the systemic redshift from \lya, that we compare to models in Sect.4. Sect.5 summarizes our findings.

\section{A sample of LAEs with known systemic redshift}
\label{s_sample}

  \begin{figure*}
    \begin{tabular}{ccc}
       \includegraphics[width=0.33\textwidth, height = 0.2\textheight]{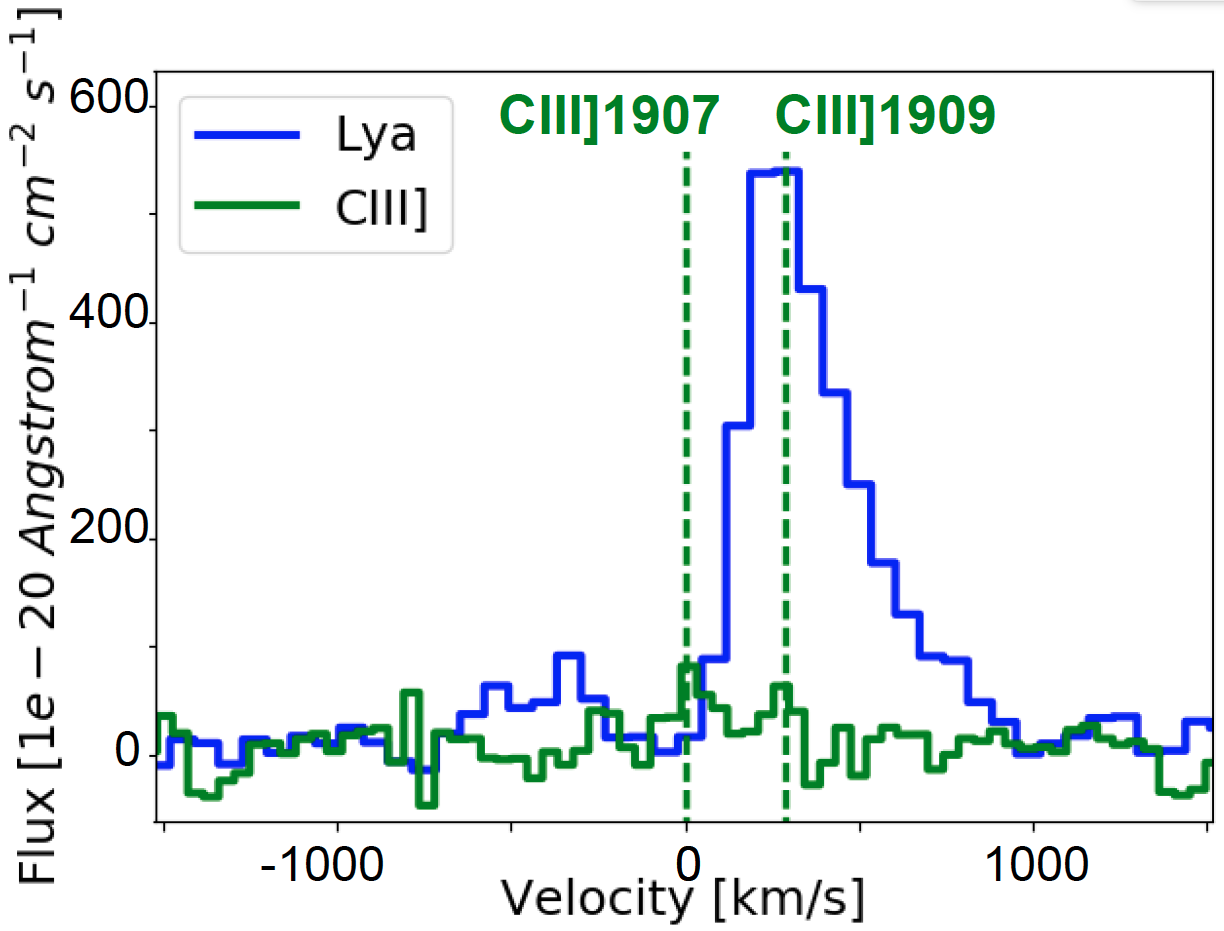}
 &
        \includegraphics[width=0.33\textwidth, height = 0.2\textheight]{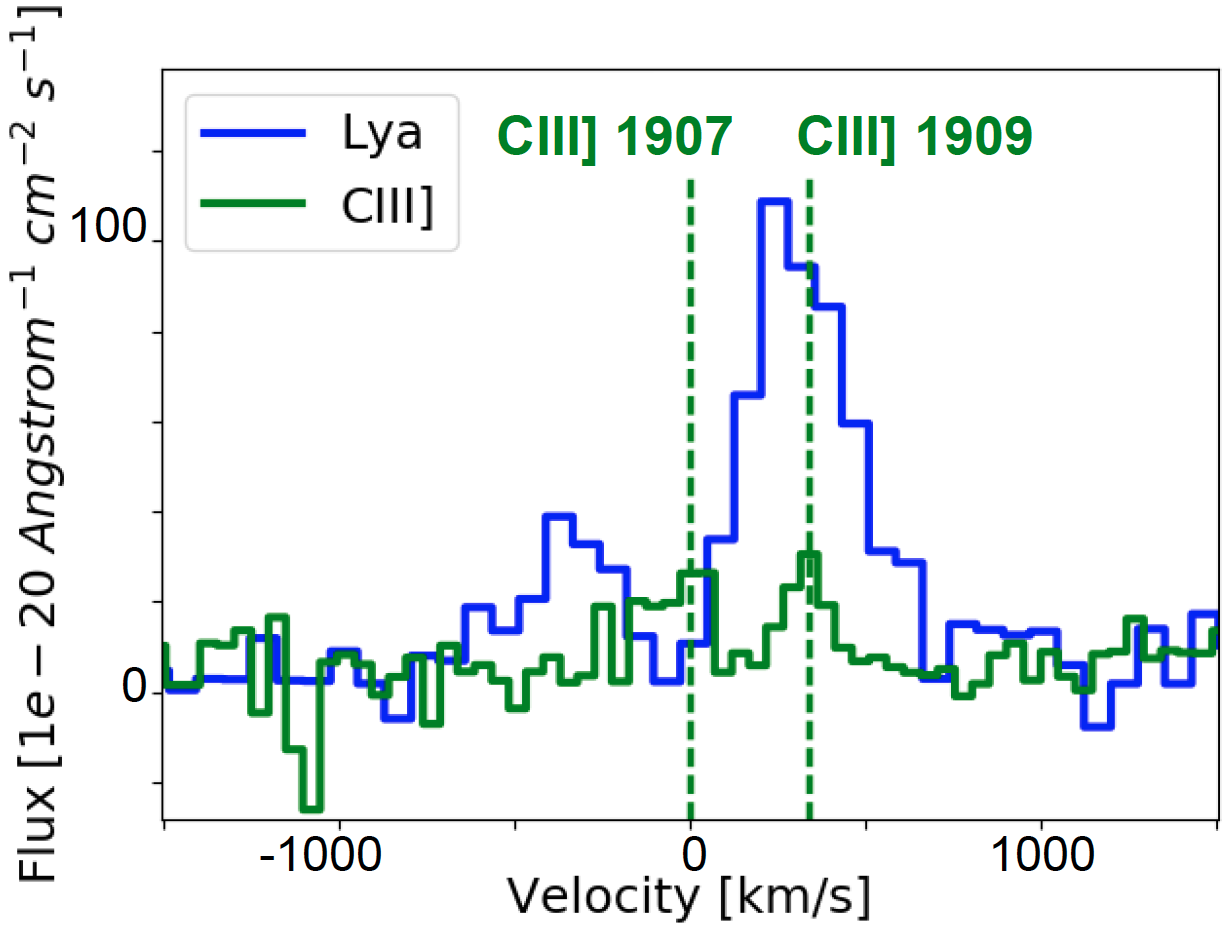} 
 &
        \includegraphics[width=0.33\textwidth, height = 0.2\textheight]{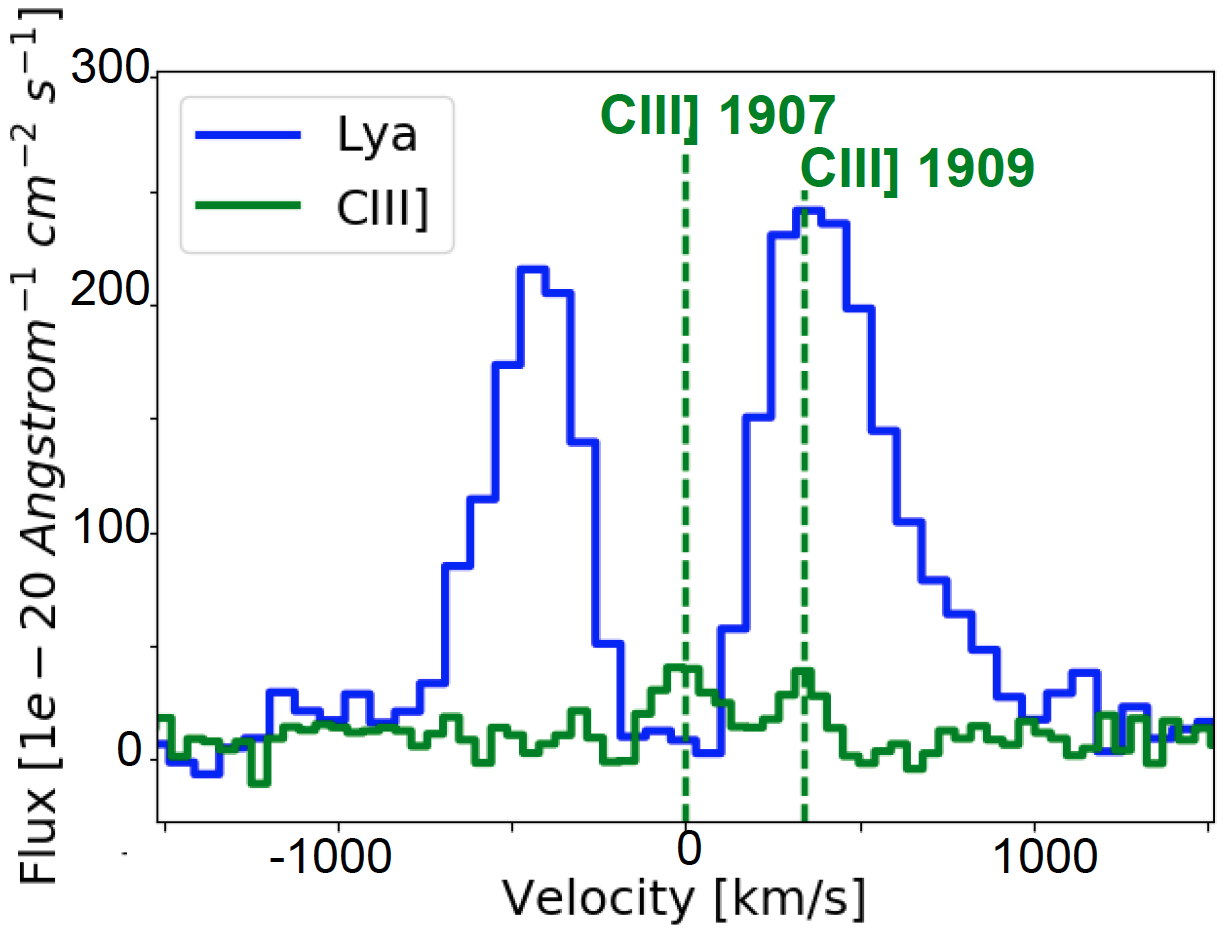} \\
        \end{tabular}
           \caption{Three examples of rest-frame spectra of LAEs with detected \ciiill\ doublet probing the systemic redshift in the MUSE-GTO observations. In each panel, the velocity shifts of the \lya\ line are shown relative to \ciii1907\AA. For all blue bump LAEs in our sample the systemic redshift falls in between the blue and red peaks of the \lya\ emission.
              }
             \label{CIII_bb}
   \end{figure*}

In order to investigate the link between the shape of the \lya\ line and the systemic redshift, we collect a diverse sample of LAEs with a precise measure of the systemic redshift. Our sample consists of high-redshift ($z > 2$) LAEs with detected \ciiill, \oiiill\ or H$\alpha\lambda6563$ emission, and low redshift ($z < 0.4$) LAEs with \lya\ observations in the UV rest-frame obtained with the Cosmic Origins Spectrograph onboard \emph{HST}, and ancillary optical spectra from the SDSS database containing several nebular emission lines from which the redshift is determined with great accuracy. We present these data in the following paragraphs. 
For each \lya\ spectrum, we measure \vpeak\ as the location of the maximum of the \lya\ flux redwards of the systemic redshift, and FWHM as the width of the part of the spectrum \emph{uncorrected for instrumental broadening} with flux above half of the maximum,  both directly on the data, without any modeling.  

\subsection{ LAEs from MUSE GTO data} 
\cite{Stark14} reported the detection of \ciiill\ emission from low mass star-forming galaxies.
When observed, this doublet is the strongest UV emission line after \lya, and, in contrast to \lya, it is an optically thin nebular line, tracing the systemic redshift\footnote{
CIII] can be used as a redshift indicator when the two components of the doublet are well resolved, i.e. when $R\sim 2000$, because their relative strength depends on the density.
} of the \lya\ production site. 
The redshift window where \lya\ and \ciii\ are both observable within the VLT/MUSE spectral range is $2.9 < z < 3.8$. MUSE is an optical Integral Field Unit (IFU) spectrograph with medium spectral resolution (from R$\sim2000$ in the blue to R$\sim4000$ in the red). 

Within several projects in the MUSE consortium using Guaranteed Time Observations \citep[GTO,][]{Bacon17, Inami17, Brinchmann17, Maseda17, Herenz17, Mahler17, Caruana18} and Science Verification (SV) or commissioning data \citep{Patricio16}, we find 13 LAEs with reliable \ciii\ detections, that is, non contaminated by sky lines and with a S/N $ > 3$. We list these objects in 
Table~\ref{table:1} \citep[see][for a systematic study of C{\sc III} emitters in the MUSE GTO data from the Hubble Ultra Deep Field]{Maseda17}. For each of these LAEs we measure the shift of the \lya\ emission compared to \ciii,\vpeak, the observed FWHM, and the separation of the \lya\ peaks for the 8 blue bump LAEs among them.

 


\subsection{High-z data from the literature}
\cite{Stark17} reported the detection of \ciii\ from one of the highest redshift \lya\ emitters ever observed ($z \sim7.730$) with \vpeak $\sim340$ \kms; the \lya\ FWHM $=360^{+90}_{-70}$ \kms\ is measured by \cite{Oesch15}. \cite{Vanzella16a} report a narrow \lya\ line observed at medium spectral resolution using VLT-Xshooter of a magnified star-forming galaxy at $z = 3.1169$, with \vpeak(\lya) $\sim100$ \kms\ and FWHM(\lya) $\sim104$ \kms. 
From \cite{Hashimoto15,Hashimoto17}, we select the 6 LAEs observed with MagE (R$\sim 4100$). Their systemic redshifts have been obtained with either H$\alpha$ or \oiii\ lines. Three of these objects are blue-bump LAEs, for which we also measure the separation of the peaks. \cite{Kulas12} reported that a significant fraction ($\sim30$\%) of their \lya\ emitting LBGs show a complex \lya\ profile, with at least one secondary peak. Blue bump objects (their Group I) represent the majority of their profiles (11 out of 18 objects). We add these 11 objects to our sample of blue bumps LAEs.

\subsection{Low-z data from the literature}

Green Pea galaxies (hereafter GPs) are LAEs in the local Universe \cite[$z \sim$ 0.1 to 0.3;][]{Jaskot14,Henry15, Verhamme17, Yang17}. The systemic redshift of these objects was compiled from the several nebular lines contained in their SDSS optical spectrum \citep[e.g ][]{Izotov11}. 
Note that the \ciii\ emission line is out of the UV spectral range probed by the available \emph{HST}-COS observations. 
For a sample of 17 GPs from \citet{Jaskot14, Henry15, Verhamme17}, we measure \vpeak\ and FWHM on the data. For 21 new GP observations, we use the \vpeak\ and FWHM values computed by \cite{Yang17} given in their Table 2. GPs nearly always exhibit blue bump \lya\ profiles \citep{Jaskot14,Henry15,Verhamme17}. For the blue bump GPs, we also measure the separation of the peaks.

\section{Deriving systemic redshift from Lyman-alpha}
\label{s_results} 

  \begin{figure*}
	  \begin{tabular}{cc}
       	\includegraphics[width=0.5\textwidth]{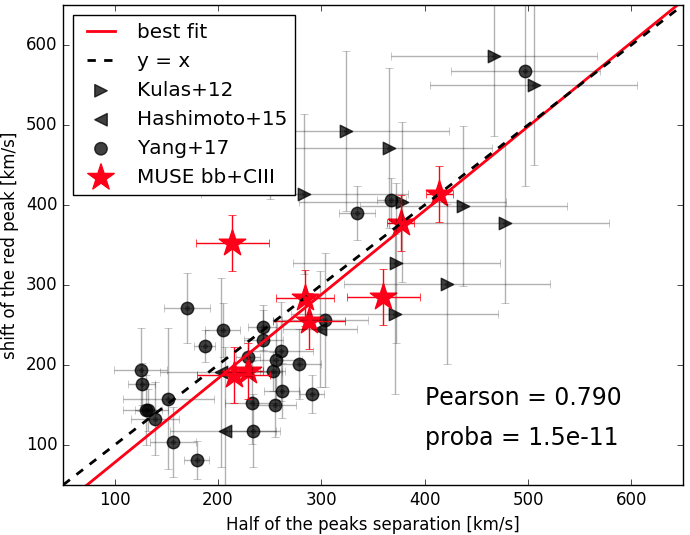} &
       	\includegraphics[width=0.5\textwidth]{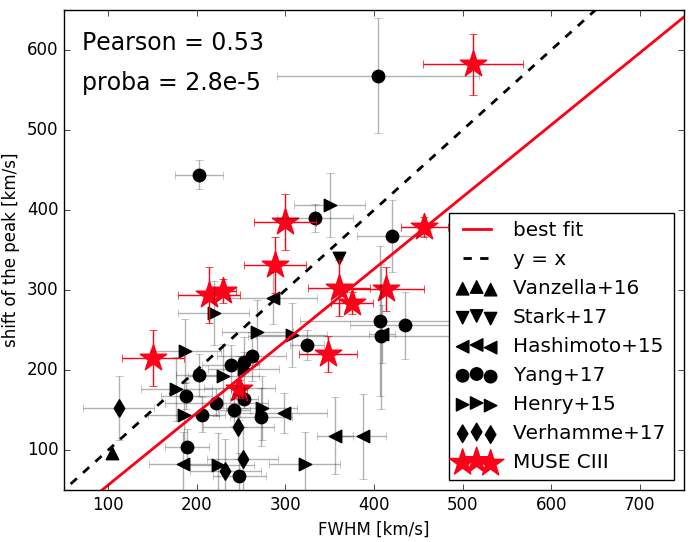} 
	\end{tabular}	
    \caption{Empirical relations to determine systemic redshift from the shape of the \lya\ emission. {\bf Left:} correlation between the shift of the \lya\ red peak, (\vpeak) and half of the separation of the peaks (\sep) for a sample of LAEs with a known systemic redshift: 7 \lya+\ciii\ emitters with blue bump \lya\ spectra from the MUSE GTO data (red stars), blue bump LAEs among the \citet{Yang17} GP sample (black dots), blue bump LAEs among the \citet{Hashimoto15} MagE sample and Group I LBGs from \citet{Kulas12} (black triangles). 
    {\bf Right:} correlation between \vpeak\ and FWHM among \lya + \ciii, H$\alpha$ or \oiii\ emitters. The black dashed line is the one-to-one relation. We checked that the correlation remains even discarding the two most upper left points. On both sides, the red curve is our best fit to the data, described by Eqs. (1) and (2). The Pearson coefficient and the probability of the null hypothesis are shown on each panel.}
             \label{results}
   \end{figure*}
   

\subsection{Method 1: systemic redshift of blue bump LAEs}

In this section, only \emph{blue bump spectra}, i.e. double peaks with a red peak higher than the blue peak, are considered. We note that, for all blue bump LAEs studied here, the systemic redshift always falls in-between the \lya\ peaks, as illustrated in Fig~\ref{CIII_bb} for three blue-bump MUSE \lya+\ciii\ emitters \citep[see also][]{Kulas12, Erb14, Yang16}. 
Fig.~\ref{results}, left panel, shows a positive empirical correlation between \vpeak\ and half of the separation of the peaks, \sep, for blue bump LAEs with known systemic redshift. 
We fit the data using the \verb+LTS_LINEFIT+ program described in \cite{Cappellari13}, which combines the Least Trimmed Squares robust technique of Rousseeuw \& van Driessen (2006) into a least-squares fitting algorithm which allows for errors in both variables and intrinsic scatter\footnote{http://www-astro.physics.ox.ac.u/$\sim$mxc/software/\#lts}. The best fit, shown by the red line on Fig.~\ref{results}, is given by:
\begin{equation}
V_{\mathrm{peak}}^{\mathrm{red}}= 1.05 (\pm 0.11) \times \Delta V_{1/2} - 12 (\pm 37){\rm km.s}^{-1} 
\end{equation}
This relation is so close to the one-to-one relation that we assume from now that the underlying "true" relation between \vpeak\ and \sep\ is one-to-one, as expected from radiation transfer modeling (see Sect.\ref{models} below). The intrinsic scatter estimated from the linear regression is $53 (\pm 9)$ \kms.

\subsection{Method 2: an empirical correlation between FWHM and systemic redshift}
\label{ss_method2}

In this section, both single and double peaked profiles are considered. The measurements are always done on the red peak, and the red peak only. In the right panel of Fig.~\ref{results} we plot \vpeak\ versus FWHM for the full sample of LAEs presented in Sect.~\ref{s_sample} (new MUSE LAEs measurements are reported in Table~\ref{table:1}). There is a correlation between \vpeak\ and FWHM although less significant than for Method 1 (see the Pearson coefficients on each panel of Fig~\ref{results}). 
We use the same method \citep{Cappellari13} to determine the empirical relation, which can be used to retrieve the systemic redshift of a galaxy: 
\begin{equation}
V_{\mathrm{peak}}^{\mathrm{red}} = 0.9 (\pm 0.14) \times {\rm FWHM(\lya)} -34 (\pm 60)\, {\rm km.s}^{-1} 
\end{equation}
This relation is also compatible with the one-to-one relation, given the uncertainties in the fit parameters. The intrinsic scatter estimated from the linear regression is $72 (\pm 12)$ \kms, slightly larger than with method 1.

\subsection{Comparison of the methods}

  \begin{figure}
       \includegraphics[width=0.5\textwidth]{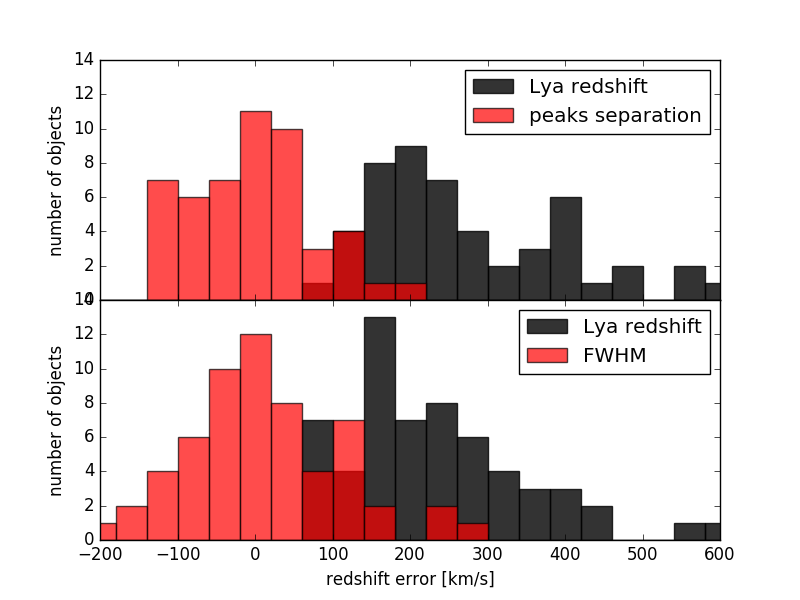} \\
    \caption{Comparison of the distributions of \lya\ redshift errors ($=z_{\lya} - z_{\rm sys}$, in black) with redshift distributions corrected with method 1 (in red, top panel) and with method 2 (in red, bottom panel). 
              }
             \label{compare}
   \end{figure}

We check that the corrected redshifts from both methods give results that are closer to the systemic redshift of the objects than the "\lya\ redshifts", i.e. taking \vpeak\ as the systemic redshift, as usually done (Fig~\ref{compare}). The standard deviation of the red histograms (corrected redshifts), reflecting both the intrinsic scatter and measurement errors, are comparable for the two methods, though slightly better for the blue bump method
. We therefore propose to use half of the separation of the peaks as a proxy for the red peak shift of blue bump LAEs, and the \lya\ FWHM for single peaked spectra\footnote{We have also tested the relation between \lya\ EWs and \vpeak, but did not find any significant correlation.}.  
They allow to recover the systemic redshift from the \lya\ line, with an uncertainty lower than $\pm100$ \kms\ from $z \approx$ 0 to 7. This suggests that the same scattering processes, linking the line shift and the line width, are at play at every redshift, and that the effect of the IGM does not erase this correlation.

\section{Discussion}
\label{s_discussion}

\subsection{Effect of the spectral resolution 
}
\label{ss_SpecRes}

These two methods to retrieve the systemic redshift of a LAE from the shape of its \lya\ profile rely on measurements of either the positions of the blue and red \lya\ emission peaks or the (red peak) FWHM. Both of these measures are affected by the spectral resolution. Although the data points presented in Sect 3. were collected from the literature and MUSE surveys and span a range of spectral resolutions from R $\sim1000$ (LRIS) to R $\sim5000$ (X-Shooter, \emph{HST}-COS), they all seem to follow the same relation. 


We investigated the effect of spectral resolution on synthetic spectra constructed from \lya\ radiation transfer simulations. Poorer spectral resolution broadens the peaks, and since \lya\ profiles are often asymmetric, it also has the effect of shifting the peak towards longer wavelengths. The latter effect is weaker than the broadening. As a consequence, the effect of spectral resolution may flatten the slope but seems not to break the correlation. 

\subsection{Comparison with models}
\label{ss_models}


  \begin{figure*}
  \begin{tabular}{cc}
       \includegraphics[width=0.5\textwidth]{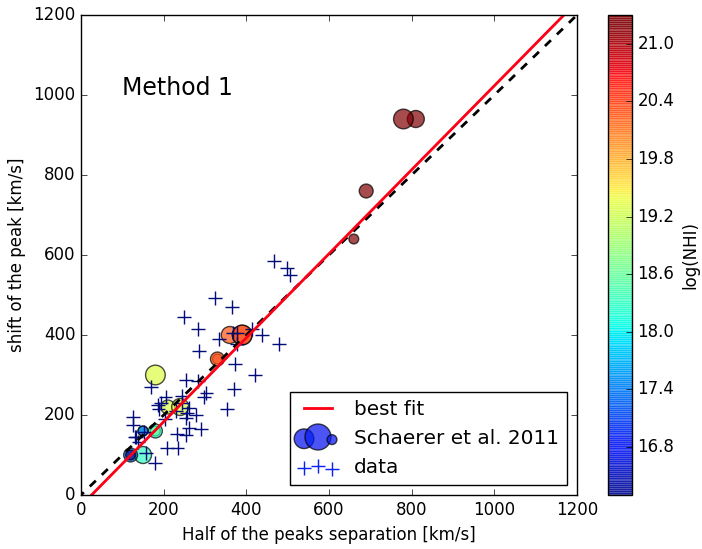} &
       \includegraphics[width=0.5\textwidth]{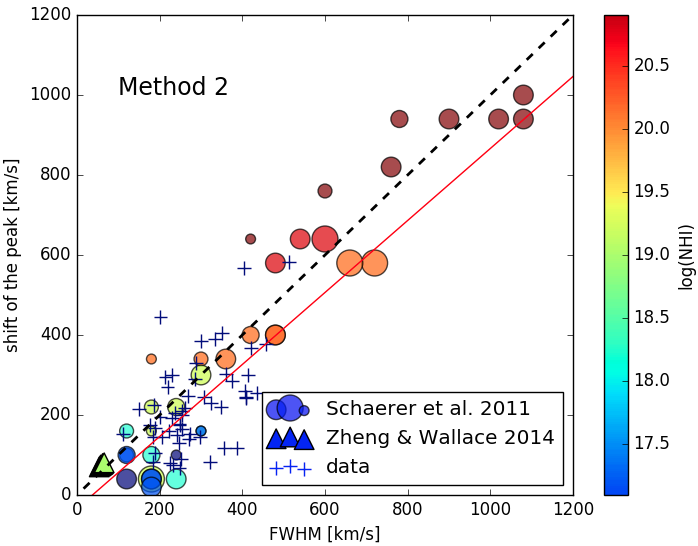} \\
	\end{tabular}       
           \caption{Points show the relationship between half of the separation of the peaks and the shift of the \lya\ line, and between the FWHM and the shift of the \lya\ line, for synthetic spectra from expanding shells, spheres or bi-conical outflows \citep{schaerer2011,Zheng14}. The trend is driven by the column density of the scattering medium, but holds for the different idealized geometries. The symbol colors scale with the column density (in cm$^{-2}$) of the shells and symbol sizes scale with the radial expansion velocity (from 0 to 400 \kms). The red line and dashed black line are identical as in  Fig~\ref{results}. 
              }
             \label{models}
   \end{figure*}

We now compare our results with numerical simulations of \lya\ radiative transfer in expanding shells performed with the MCLya code \citep{schaerer2011, Verhamme06}. 
These models describe in a simple, idealized, way the propagation of \lya\ photons emitted in HII regions through gas outflows which seem ubiquitous in star-forming galaxies, especially at high redshift \citep{shapley03, steidel2010,Hashimoto15}. 
Assuming a central point-source surrounded by an expanding shell of gas with varying HI column density ($N_{\rm HI}$), speed ($V_{\rm exp}$), dust opacity ($\tau_{\rm d}$) and temperature (described by the Doppler parameter $b \propto \sqrt{T}$), shell models have proven very successful in reproducing a large diversity of \lya\ line profiles.  Here, we use simulations with different intrinsic Gaussian line widths ($\sigma_i$) and various shell parameter values ($N_{\rm HI}$, $V_{\rm  exp}$, $\tau_{\rm d}$, $b$), degraded to mimic the MUSE spectral resolution.
We measure FWHM, \vpeak\ and \sep\ the same way as for the data.  

We compare the observed correlation between \vpeak\ and the separation between the peaks of blue-bump LAEs (\sep) with results from models that produce double-peak profiles (Fig.~\ref{models}, left panel). Predictions from expanding shell models lie very close to the one-to-one relation and reproduce nicely the observed properties of the \lya\ lines.  Objects with increasing \vpeak\ and \sep\ correspond to expanding shells with larger HI column densities. This echoes the analytical solutions for \lya\ RT in static homogeneous media \citep{neufeld90,dijkstra06} that yield profiles with symmetric peaks around the line centre, whose positions are primarily set by the HI opacity and correspond to \vpeak $\propto \tau_{\rm HI}^{1/3}$.  

As shown in Fig.~\ref{models} (right panel), the correlation between the shift of the red peak and the FWHM of the \lya\ line naturally arises from scattering processes. The slope predicted by the models is close to one whereas the relation derived from observations in Section 3 is shallower ($\approx 0.9$; red curve in the right panel of Figs.~\ref{results},\ref{models}). However, it is worth pointing out that we explore a much larger range of FWHMs in the right panel of Fig.~\ref{models} (from 0 to 1200 \kms) compared to Fig.~\ref{results} where observed FWHMs vary from 214 to 512 \kms. For FWHM values less than 600 \kms, the model predictions lie close to the FWHM-\vpeak\ relation derived in Section 3.
Although the exact location of each simulated object in the FWHM-\vpeak\ plane seems to depend on each parameter, we see that models with higher HI column
densities lead to broader lines and larger shifts of the peak (color-coded circles). A similar trend is found by \citet{Zheng14} who performed \lya\ radiation transfer simulations in anisotropic configurations (bipolar outflows) and inhomogeneous media (i.e. HI distributions with velocity or density gradients). Overall, this may suggest that the FWHM-\vpeak\ correlation holds regardless of the assumed geometry and kinematics of the outflows, and that the HI opacity of the ISM and/or the medium surrounding galaxies (i.e. the CGM) is the main driver that shapes the observed \lya\ line profiles.


%
 
\section{Conclusions}
\label{s_ccl}

The recent increase in the number of LAEs with detected nebular lines allows to calibrate empirical methods to retrieve the systemic redshift from the shape of the \lya\ line.
In addition to measurements from the literature, we report 13 new detections from several MUSE GTO programs.
We searched for \lya+\ciii\ emitters in the MUSE-Deep survey (Bacon et al. 2017), 
behind $z \sim 0.7$ galaxy groups (Contini et al, in prep), and lensed by three clusters \citep[SMACSJ2031.8-4036 in][AS1063, MACS0416 in Richard et al, in prep]{Patricio16}.

We find a robust correlation between the shift of the \lya\ peak with respect to systemic redshift (\vpeak) and half of the separation of the peaks (\sep) for LAEs with blue bump spectra. The intrinsic scatter around the relation is $\pm 53$ \kms.
We also find a correlation between the shift of the \lya\ peak with respect to systemic redshift (\vpeak) and its width at half-maximum (FWHM), for LAEs with known systemic redshift. The intrinsic scatter is of the same order ($\pm 73$ \kms). These two relations have been approximated by linear fitting formulas as given in Eq (1) and (2). These formulae have been derived for data with spectral resolution $1000 < R < 5000$, they should be used on data with similar spectral resolution.

The relative redshift error if estimated from \lya\ with \vpeak$=300$ \kms\ is ($\Delta$z /z)(\lya)=((1+z)$\times$(\vpeak / c))/z $\sim10^{-3}$ at $z=3$. The two methods presented in this letter can therefore help reduce systematic errors on distance measures. This is of great importance for redshift surveys at $z \gtrsim 3$, where spectroscopic redshifts often rely on the \lya\ emission line. Futures observations with better spectral resolution should allow to refine the proposed relations.

\section*{Acknowledgements}
We thank the anonymous referee for her/his helpful report.
AV is supported by a Marie Heim V\"ogtlin fellowship of the Swiss National Foundation. 
TG is grateful to the LABEX Lyon Institute of Origins (ANR-10-LABX-0066) of the Université de Lyon for its financial support within the program "Investissements d'Avenir" (ANR-11-IDEX-0007) of the French government operated by the National Research Agency (ANR).
TC, EV, JZ acknowledge support of the ANR FOGHAR (ANR-13-BS05-0010-02), the OCEVU Labex (ANR-11- LABX-0060) and the A*MIDEX project (ANR-11- IDEX-0001-02) funded by the “Investissements d’Avenir” French government program managed by the ANR.
RB and FL acknowledges support from the ERC advanced grant 339659-MUSICOS. 
JR and VP acknowledge support from the ERC starting grant 336736-CALENDS.
RAM acknowledges support by the Swiss National Science Foundation. 
JS acknowledges support from the ERC grant 278594-GasAroundGalaxies.
JB acknowledges support by Funda{\c c}{\~a}o para a Ci{\^e}ncia e a Tecnologia (FCT) through national funds (UID/FIS/04434/2013) and by FEDER through COMPETE2020 (POCI-01-0145-FEDER-007672) and Investigador FCT contract IF/01654/2014/CP1215/CT0003.



\bibliographystyle{mnras}
\bibliography{refs} 




\appendix

\begin{table*}
\caption{
MUSE \lya+\ciii\ emitters. The 6th column indicates the separation of the peaks (i.e. $2\times$\sep, in \kms) for blue bump LAEs, and is left empty for single-peaked profiles. 
a: Patricio et al. 2016; b: Richard et al. 2018 in prep; c: Bacon et al. 2017, Inami et al. 2017, Maseda et al. 2017; d: Contini et al. 2018 in prep.
}
\label{table:1}      
\centering                          
\begin{tabular}{|c c c c c c c c c|}        
\hline                 
 ID & RA & DEC & EW [\AA] &\vpeak [\kms] & FWHM [\kms] & $\Delta$V [\kms]& z$_\textrm{sys, CIII]}$ &  observations \\    
\hline \hline                        
sys 1$^{a}$  & 307.97040 & -40.625694 &   32 & $176\pm 11$ & $248\pm 9$ &  -- & 3.5062 & commissioning \\      
mul 11$^{b}$ & 342.175042 & -44.541031 & 222 & $215\pm 35$ & $150\pm 35$ & $375\pm 35$ & 3.1163 & AS1063  \\ 
mul 14$^{b}$ & 342.178833 & -44.535869 &  29 & $385\pm 35$ & $300\pm 35$ & -- & 3.1150 & AS1063 \\
sys  44$^{b}$& 64.0415559 & -24.0599916 & 57 & $303\pm 35$ & $360\pm 35$ & $570 \pm 35$ & 3.2886 & MACS0416 \\ 
sys 132$^{b}$& 64.0400838 & -24.0667408 & 62 & $331\pm 35$ & $288\pm 35$ & $510 \pm 35$ & 3.2882 & MACS0416 \\ 
 106$^{c}$  & 53.163726  & -27.7790755 & 72 & $379\pm 13$ & $414\pm 13$ & $828 \pm 35$ & 3.2767 & \udft \\
 118$^{c}$  & 53.157088  & -27.7802688 & 65 & $301\pm 28$ & $284\pm 28$ & $568 \pm 35$ & 3.0173 & \udft \\
1180$^{c}$  & 53.195735  & -27.7827171 & 80 & $220\pm 23$ & $348\pm 32$ & -- & 3.3228 & \mosaic \\
6298$^{c}$  & 53.169249  & -27.7812550 & 83 & $582\pm 38$ & $512\pm 56$ & -- & 3.1287 & \udft \\
6666$^{c}$  & 53.159576  & -27.7767193 & 52 & $284\pm 13$ & $377\pm 11$ & $754 \pm 35$ & 3.4349 & \udft \\
  50$^{d}$  & 150.149656 & 2.061272 & 50 & $431\pm 42$ & $268\pm 39$ & -- & 3.8237 & GR30   \\
  48$^{d}$  & 149.852989 & 2.488099 & 68 & $294\pm 35$ & $214\pm 35$ & $705 \pm 35.$ & 3.3280 & GR34 \\
 102$^{d}$  & 150.050268 & 2.600025 & 76 &$299\pm 15$ & $229\pm 15$ & $385 \pm 35.$ & 3.0400 & GR84 \\
\hline                                   
\end{tabular}
\end{table*}




\bsp	
\label{lastpage}
\end{document}